\documentstyle[prc,multicol,epsf,aps]{revtex} 
\begin{document} 
\draft 
\title{Shell model calculation of the $\beta^-$ and $\beta^+$ partial
  halflifes of $^{54}$Mn and other unique second forbidden $\beta$
  decays}

\author{Gabriel Mart\'{\i}nez-Pinedo$^{1,2}$ and Petr Vogel$^3$ }

\address{$^1$Institute of Physics and Astronomy, University of
  {\AA}rhus, DK-8000 {\AA}rhus, Denmark}

\address{$^2$Kellogg Radiation Laboratory 106-38, California Institute
  of Technology, Pasadena, CA 91125, USA} 

\address{$^3$Department of Physics 161-33, California Institute
of Technology, Pasadena, CA 91125, USA}

\date{\today} 

\maketitle

\begin{abstract}
  The nucleus $^{54}$Mn has been observed in cosmic rays.  In
  astrophysical environments it is fully stripped of its atomic
  electrons and its decay is dominated by the $\beta^-$ branch to the
  $^{54}$Fe ground state.  Application of $^{54}$Mn based chronometer
  to study the confinement of the iron group cosmic rays requires
  knowledge of the corresponding halflife, but its measurement is
  impossible at the present time. However, the branching ratio for the
  related $\beta^+$ decay of $^{54}$Mn was determined recently. We use
  the shell model with only a minimal truncation and calculate both
  $\beta^+$ and $\beta^-$ decay rates of $^{54}$Mn. Good agreement for
  the $\beta^+$ branch suggests that the calculated partial halflife
  of the $\beta^-$ decay, $(4.94\pm 0.06)\times 10^5$ years, should be
  reliable.  However, this halflife is noticeably shorter than the
  range 1-$2\times 10^6$~y indicated by the fit based on the $^{54}$Mn
  abundance in cosmic rays. We also evaluate other known unique second
  forbidden $\beta$ decays from the nuclear $p$ and $sd$ shells
  ($^{10}$Be, $^{22}$Na, and two decay branches of $^{26}$Al) and show
  that the shell model can describe them with reasonable accuracy as
  well.
\end{abstract} 

\begin{multicols}{2}

\section{Introduction}

\begin{figure}
  \begin{center}
    \leavevmode 
    \epsfxsize=0.44\textwidth 
    \epsffile{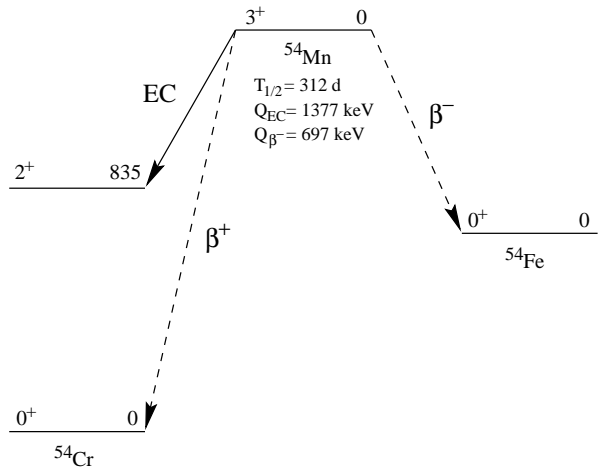}
    \caption{Decay scheme of $^{54}$Mn.}
    \label{fig:decay}
  \end{center}
\end{figure}

The nucleus $^{54}$Mn decays in the laboratory dominantly by electron
capture to the $2^+$ state in $^{54}$Cr with the halflife of 312 days.
However, as a component of cosmic rays, $^{54}$Mn will be fully
stripped of its atomic electrons, and this mode of decay is therefore
impossible. The $^{54}$Mn nuclei were in fact detected in cosmic rays
using the Ulysses spacecraft~\cite{Ulysses1,Ulysses2}. They offer an
attractive possibility to use their measured abundance as a
chronometer for the iron group nuclei (Sc - Ni) in cosmic rays in
analogy to the chronometers based on the abundances of other long
lived isotopes ($^{10}$Be, $^{26}$Al, and $^{36}$Cl). With them one
can, in turn, determine the mean density of interstellar matter, a
quantity of considerable interest. The use of the long lived nuclei as
cosmic ray chronometers is reviewed in Ref.~\cite{Simpson}. The
importance of $^{54}$Mn for the understanding of propagation of the
iron group nuclei that are products of explosive nuclear burning have
been stressed in Refs.~\cite{Grove,Leske}.  For this program to
succeed, however, one must know the halflife of the stripped $^{54}$Mn
($I^{\pi} = 3^+$). The decay scheme of $^{54}$Mn is shown in
Fig.~\ref{fig:decay}; the dashed lines indicate the decay paths of the
stripped $^{54}$Mn.

In two recent difficult and elegant experiments the very small
branching ratio for the $\beta^+$ decay to the ground state of
$^{54}$Cr has been measured: $(2.2 \pm 0.9) \times 10^{-9}$~\cite{b+1}
and $(1.20 \pm 0.26) \times 10^{-9}$~\cite{b+2}. By taking the weighed
mean of these values we extract the averaged branching ratio of $(1.28
\pm 0.25) \times 10^{-9}$.  Combining it with the known halflife for
$^{54}$Mn of 312.3(4)~d~\cite{nds54}, it corresponds to an
experimental partial $\beta^+$ halflife of $(6.7 \pm 1.3) \times 10^8$
years.  As explained in~\cite{Ulysses1,b+1,b+2} one expects, however,
that the decay of the fully stripped $^{54}$Mn will be dominated by
the at present unobservable $\beta^-$ decay to the $^{54}$Fe ground
state.  Previously, the partial $\beta^-$ halflife was estimated
assuming that the $\beta^-$ and $\beta^+$ form factors are identical.
Very recently, in Ref. \cite{b+2}, the ratio of the $\beta^-$ and
$\beta^+$ form factors was calculated using a very truncated
shell-model and extending it by comparison with similar calculations
in the $sd$-shell. The estimated $\beta^-$ halflife is $(6.3 \pm 1.3)
\times 10^5$~y \cite{b+2}.  In this work we will use the state of the
art shell model and evaluate not only the $EC$ decay rate of the
normal $^{54}$Mn, but also both decay branches of the unique second
forbidden transitions $^{54}$Mn$(3^+) \rightarrow {}^{54}$Cr$(0^+)$
and $^{54}$Mn$(3^+) \rightarrow {}^{54}$Fe$(0^+)$.  By comparing the
calculated $\beta^+$ decay halflife (or branching ratio) to the
measured one we hope to judge the reliability of the calculation. We
then proceed to calculate the halflife of the unknown $\beta^-$ decay.

The decays of stripped $^{54}$Mn are unique second forbidden
transitions which depend on a single nuclear form factor (matrix
element).  Halflives of several such decays in the $p$ shell
($^{10}$Be) and $sd$ shell ($^{22}$Na, and two decay branches of
$^{26}$Al) are known and have been compared to the nuclear shell model
predictions in Ref.~\cite{War}. For the $sd$ shell nuclei, however,
only calculations in a severely truncated space were performed
in~\cite{War}.  Since that time computation techniques and programming
skills have improved considerably.  Thus, in order to further test our
ability to describe this kind of weak decays, we repeat the
analysis~\cite{War}, using the exact shell model calculations without
truncation. At the same time the availability of new (and different)
experimental data for the $^{10}$Be~\cite{be10exp} and
$^{26}$Al~\cite{al26exp} decays make necessary a new comparison
between experiment and calculations.

In order to evaluate the decay rate we use the formulation
of~\cite{tables}.  The number of particles with momentum $p$ emitted
per unit time is:
\begin{equation}
  N(p_e) dp_e = \frac{g^2}{2 \pi^3} p_e^2 p_{\nu}^2 F(Z,W_e) C(W_e)
  dp_e,
\end{equation}
where $g$ is the weak coupling constant, $p_e$ and $W_e$ are electron
(or positron) momentum and energy, $p_{\nu}$ is the neutrino (or
antineutrino) momentum, and $Z$ is the atomic number of the daughter
nucleus. All momenta and energies are in units where the electron mass
is unity.  For the Fermi function $F(Z,W_e)$ we use the tabulated
values, and the shape factor $C(W_e)$ for the case of the unique
second forbidden transitions is of the form
\begin{equation}
  C(W_e) = \frac{R^4}{15^2} \left| ^AF_{321}^{(0)} \right|^2 \left[
    p_\nu^4 + \frac{10}{3} \lambda_2 p_\nu^2 p_e^2 + \lambda_3 p_e^4
  \right].
\end{equation}
The nuclear form factor, in turn, is defined as
\begin{equation}
  ^AF_{321}^{(0)} = g_A \sqrt{\frac{4\pi}{2J_i + 1}} \frac{\langle f
    || r^2 [\bbox{Y}_2 \times \bbox{\sigma}]^{[3]}\bbox{t}_\pm || i
    \rangle}{R^2},
    \label{eq:formfactor}
\end{equation}
where $i$ denotes the initial state and $f$ the final one; the matrix
element is reduced with respect to the spin space only (Racah
convention~\cite{edmons}); $\pm$ refers to $\beta^\pm$ decay;
$\bbox{t}_\pm = (\bbox{\tau}_x \pm i \bbox{\tau}_y)/2$, with
$\bbox{t}_+ p = n$; $g_A = -1.2599 \pm 0.0025$~\cite{towhar}; and $R$
is the nuclear radius (the final expression for $C(W_e)$ is obviously
independent of $R$).

The functions $\lambda_2$ and $\lambda_3$ are tabulated
in~\cite{tables}.  Integrating the rate formula up to the spectrum
endpoint we obtain the expression for $1/\tau$ and, respectively, for
the halflife ($T_{1/2} = \ln(2) \tau$) in terms of the nuclear form
factor squared. (For the stripped atoms we correct the endpoint energy
accordingly.). For the quantity $2 \pi^3(\ln 2)/g^2$ we use the value
$6146\pm 6$~s~\cite{towhar}.  Note the usual $f t$ value, commonly
used to characterize a decay, uses the integrated phase space factor
$f$ of Eq. (1), however without the constant $g^2/(2 \pi^3 15^2)$) and
the radius factor $R^4$.

\section{Shell model calculations}

\begin{figure}
  \begin{center}
    \leavevmode 
    \epsfxsize=0.45\textwidth 
    \epsffile{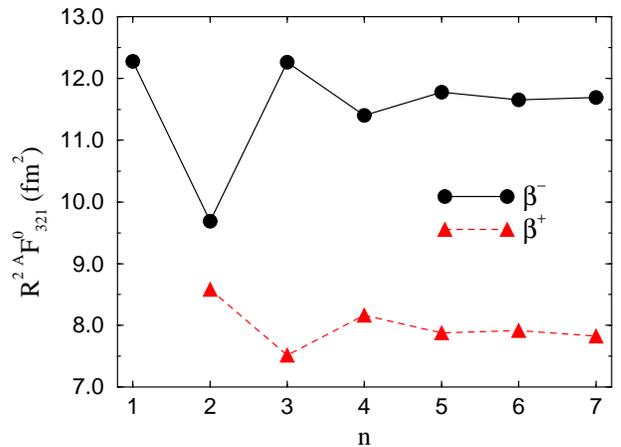}
    \caption{Evolution of the matrix element $R^2\,^AF^0_{321}$ with
      the truncation level $n$. The form factor has been evaluated
      using harmonic oscillator wave functions with $b=1.99$~fm}.
    \label{fig:Fn}
  \end{center}
\end{figure}

In our calculation we consider an inert core of $^{40}$Ca with the 14
remaining nucleons distributed throughout the $pf$-shell. We use
KB3~\cite{KB3} as the residual interaction with the single particle
energies taken from the $^{41}$Ca experimental spectrum. The
Hamiltonian is diagonalized with the code {\sc antoine}~\cite{antoine}
using the Lanczos method. It is not yet possible to perform a full
$pf$-shell calculation but we can come fairly close. Let's denote by
$f$ the $f_{7/2}$ orbit and by $r$ the rest of the $pf$ shell
($p_{3/2}$, $p_{1/2}$ and $f_{5/2}$).  From the calculations of
reference~\cite{iron} we know that one can get a good approximation to
the results in the full $pf$-shell calculation when one considers the
evolution of a given quantity as the number of particles, $n$, allowed
to occupy the $r$-orbits, increases. We have extended the calculations
of the previous reference allowing up to a maximum of $n=7$ particles
in the $r$-orbits. The $m$-scheme dimensions for this calculation are:
17,136,878 for $^{54}$Cr; 49,302,582 for $^{54}$Mn; 91,848,462 for
$^{54}$Fe. Figure~\ref{fig:Fn} shows the variation of the nuclear
form-factor for the $\beta^+$ and $\beta^-$ transitions as a function
of the truncation level $n$. We have used the harmonic oscillator wave
functions with $b = 1.99$~fm. From Figure~\ref{fig:Fn} it is clear
that our calculation has already converged for the $n=5$ truncation.
It is also obvious that it would be inappropriate to use only the
lowest order corrections ($n=2$ for $\beta^-$ and $n=3$ for
$\beta^+$).

To see how well the calculated wave function reproduces basic
characteristics of the ground state of the odd-odd nucleus $^{54}$Mn,
note that the electric quadrupole moment is calculated to be
$Q=34$~$e$~fm$^2$ (with effective charges $e_\pi=1.5$ and $e_\nu
=0.5$), while the experimental value is $Q = 33\pm3$~$e$~fm$^2$.  The
magnetic moment, calculated with the free nucleon gyromagnetic factors
is $\mu = 2.78~\mu_N$, while the experimental value is $\mu =
3.2819\pm 0.0013~\mu_N$. We have also calculated the $\log ft$ value
for the Gamow-Teller electron capture transition to the $2^+$ state in
$^{54}$Cr. The calculated $\log ft = 6.14$, where we have used the
usual quenching factor of 0.76, is in good agreement with experimental
value of 6.2. Note, however, that quenching of this allowed
Gamow-Teller matrix element is needed to achieve the agreement with
the experimental rate. (See \cite{Quench} and references therein for
the problem of the GT strength quenching.)

{}From Eq.~(\ref{eq:formfactor}) we know that the single particle
matrix elements needed for the evaluation of the form factor involve
the expectation value of $r^2$ between the single-particle radial wave
functions. In the evaluation of this quantity we have followed two
approximations. First, we consider harmonic oscillator wave functions.
In this case all matrix elements are proportional to the square of the
length parameter $b$. We use the prescription of
reference~\cite{towner} to determine $b$ from the experimental charge
radius $\langle r^2 \rangle^{1/2}_{\text{ch}}$~\cite{radius} of the
parent nucleus; this leads to $b=1.99$~fm in $^{54}$Mn.  Second, we
consider Woods-Saxon radial wave functions.  They have been obtained
using the potential well that includes spin-orbit and Coulomb
terms~\cite{bm}.  The radial parameter of the well has been adjusted
to reproduce the experimental charge radius. The values of the form
factor and halflives obtained using both methods are listed in
Table~\ref{tab:results}.

We have also evaluated the other known unique second-forbidden beta
decays: $^{10}$Be$(\beta^-)^{10}$B, $^{22}$Na$(\beta^+)^{22}$Ne and
$^{26}$Al$(\beta^+)^{26}$Mg. For the $sd$-shell nuclei we consider an
inert core of $^{16}$O and the $sd$-shell as the valence space. These
transitions were previously computed in Ref.~\cite{War} using a
truncated shell-model calculation.  In our case, without truncating
the $sd$ shell space, we use the Wildenthal USD effective
interaction~\cite{wilden} and determine the radial parameters
following the procedure outlined in the previous paragraph. In the
harmonic oscillator approximation we use $b=1.78$~fm for $^{22}$Na and
$b=1.81$~fm for $^{26}$Al. Table~\ref{tab:results} contains the
results of our calculations. For the decay of $^{10}$Be we reproduce
the results of reference~\cite{War}, however our determined $b$
parameter, 1.75~fm, is slightly larger than the one used before
($b=1.68$~fm). The new experimental value for the
halflife~\cite{be10exp} nicely agrees with the computed one.

\section{Results and discussion}

The top row of Table~\ref{tab:results} shows that our shell model
result agrees with the measured halflife of the $\beta^+$ decay
$^{54}$Mn $\rightarrow$ $^{54}$Cr within errors, without quenching of
the corresponding form factor. The calculated $\beta^-$ halflife,
$(4.94 \pm 0.06) \times 10^5$~y (if we arbitrarily take the average
value between HO and Wood-Saxon calculations), is noticeably shorter
than the range expected in Refs.~\cite{Ulysses1,Ulysses2} (1-$2\times
10^6$~y) based on the experimental abundance of $^{54}$Mn in cosmic
rays and the model of the cosmic ray confinement.

Our calculation suggests that the form factor for the $\beta^-$ decay,
11.7~fm$^2$, is larger than the form factor for the $\beta^+$ decay,
7.8~fm$^2$.  We can offer some intuitive, albeit very crude,
understanding of this difference.  Let us treat the three nuclei in
the extreme single particle model. Also, instead of the actual
transitions connecting the odd-odd nucleus $^{54}$Mn with the
corresponding even-even ground state, let us consider the transitions
from the seniority zero even-even nuclei to the odd-odd one. The Cr
$\rightarrow$ Mn transition would then involve $(\pi f_{7/2})^4(\nu
p_{3/2})^2$ $\rightarrow (\pi f_{7/2})^5(\nu p_{3/2})$, changing a
$p_{3/2}$ neutron into a $f_{7/2}$ proton.  In contrast, the Fe
$\rightarrow$ Mn would involve $(\pi f_{7/2})^6 \rightarrow (\pi
f_{7/2})^5(\nu p_{3/2})$, changing a $f_{7/2}$ proton into a $p_{3/2}$
neutron.  Using the above naive assignments, we are led to the
conclusion that the $\beta^-$ form factor should be about $\sqrt{3}$
times larger than the $\beta^+$ form factor.  Even though the detailed
shell model results are not fully determined by the indicated single
particle transitions (but they are the largest ones), the overall
scaling factor emerges.

Among the $sd$ shell transitions in Table~\ref{tab:results}, the
transition to the 1.8~MeV state in $^{26}$Mg agrees perfectly with the
experiment, while the calculated form factors for the other two are
somewhat larger, by a factor of about 1.5, than the experimental
value. For $^{10}$Be decay the calculated form factor is also a bit
larger. We cannot, therefore, draw any conclusion about the necessity
of quenching in the case of the unique second forbidden transitions.
While the lighter $p$ and $sd$ shell nuclei contain perhaps a hint
that quenching is needed, it would obviously spoil the agreement in
the case of the $\beta^+$ branch of the $^{54}$Mn decay.

In conclusion: Our shell model calculations reproduce the experimental
halflives of the unique second forbidden beta decays within a factor
of less than two. No clear evidence for the quenching of the
corresponding form factors emerges. For the stripped $^{54}$Mn decays,
the shell model describes the $\beta^+$ branch within errors. It
predicts that the form factor for the $\beta^-$ decay is larger than
the one for the $\beta^+$ decay. The calculated $\beta^-$ halflife
(and therefore also the total halflife) is noticeably shorter than the
range based on the observation of $^{54}$Mn in cosmic rays. This
conflict, albeit relatively mild, makes attempts to determine the
branching ratio for the $\beta^-$ decay experimentally even more
compelling.

\acknowledgments

P.V. is supported in part by the U. S. Department of Energy under
grant No. DE-FG03-88ER-40397. G.M.P. is supported in part by the
DGICyES (Spain). Computational resources were provided by the Center
for Advanced Computational Research at Caltech.

\narrowtext

\end{multicols}

\widetext

\begin{table}
  \begin{center}
    \squeezetable
    \caption{Form factors and half-lives for the $\beta^+$ and
      $\beta^-$ unique second-forbidden transitions. }
    \label{tab:results}
    \renewcommand{\arraystretch}{1.2}
    \begin{tabular}{lcccccc}
      & \multicolumn{3}{c}{$R^2\,{}^AF^0_{321}$ (fm$^2$)} &
      \multicolumn{3}{c}{halflife (years)} \\ \cline{2-4} \cline{5-7}
      & HO & Woods-Saxon & Exp. & HO & Woods-Saxon & Exp. \\ \hline
      $^{54}$Mn$(\beta^+)^{54}$Cr & 7.82 & 7.76 & $7.1\pm0.7$ & $5.55
      \times 10^8$ & $5.64 \times 10^8$ & $(6.7 \pm 1.3)\times
      10^8$ \\
      $^{54}$Mn$(\beta^-)^{54}$Fe & 11.7 & 11.6 & &
      $4.89 \times 10^5$ & $4.98 \times 10^5$ & \\
      $^{22}$Na$(\beta^+)^{22}$Ne & 9.24 & 9.78 & $6.0\pm 0.8$ &
      $2.04 \times 10^3$ & $1.87 \times 10^3$ & $(4.8 \pm 1.3 )\times
      10^3$ \\
      $^{26}$Al$(\beta^+)^{26}$Mg\tablenotemark[1] & 2.44 & 2.78 &
      $2.38\pm0.05$ & $8.64 \times 10^5$ & $6.65\times 10^5$ &
      $(9.1\pm 0.4)\times 10^5$ \\
      $^{26}$Al(EC)$^{26}$Mg\tablenotemark[1] & 2.44 & 2.78 &
      $2.39\pm0.05$ & $4.58\times 10^6$ & $3.52\times 10^6$ &
      $(4.8\pm0.2)\times 10^6$ \\
      $^{26}$Al(EC)$^{26}$Mg\tablenotemark[2] & 12.6 & 13.8 &
      $8.8\pm0.5$ & $1.43\times 10^7$ & $9.44 \times 10^6$ &
      $(2.7\pm 0.3)\times 10^7$ \\
      $^{10}$Be$(\beta^-)^{10}$B & 23.1 & 23.3 & $20.4\pm 0.4$ &
      $1.18 \times 10^6$ & $1.16 \times 10^6$ & $(1.51\pm 0.06)\times
      10^6$
    \end{tabular}
    \tablenotetext[1]{The first-excited state at 1.809~MeV in
    $^{26}$Mg} 
    \tablenotetext[2]{The second-excited state at 2.938~MeV in
    $^{26}$Mg} 
  \end{center}
\end{table}

\end{document}